\documentclass[aps,twocolumn,pre,showpacs]{revtex4}
\usepackage{amssymb}
\usepackage{graphicx}
\usepackage{amssymb}
\usepackage{wasysym}
\usepackage{gensymb}

\begin{document}

\title{Experiments on the morphology of icicles}

\author{Antony Szu-Han Chen and Stephen W. Morris}

\affiliation{Department of Physics, University of Toronto, 60 St. George St., Toronto, Ontario, Canada M5S 1A7}

\date{\today}

\begin{abstract}
Icicles form when cool water drips from an overhanging support under ambient conditions which are below freezing.  Ice growth is controlled by the removal of latent heat, which is transferred into the surrounding air \emph{via} a thin film of water flowing over the ice surface.  We describe laboratory experiments in which icicles were grown under controlled conditions.  We used image analysis to probe the evolution of the icicle shape under various conditions.  A recent asymptotic theory suggests that, overall, icicles converge to self-similar shapes which are predicted to be attractors.  On the other hand, stability theory predicts that the ice-water interface can become unstable to form ripple patterns on the icicle surface.  Our experimental results show that the predicted self-similar profile is only found in certain cases, and that icicles can also exhibit unpredicted  non-uniformities such as branching near the tip.  We find that pure water icicles are more likely to be self-similar than those grown from tap water.   Ripples, which are also deviations from the self-similar profile, were observed to climb upward during icicle growth.
\end{abstract}

\pacs{47.54.-r,81.10.-h,47.20.Hw}

\maketitle

\begin{figure}
\includegraphics[width=8.8cm]{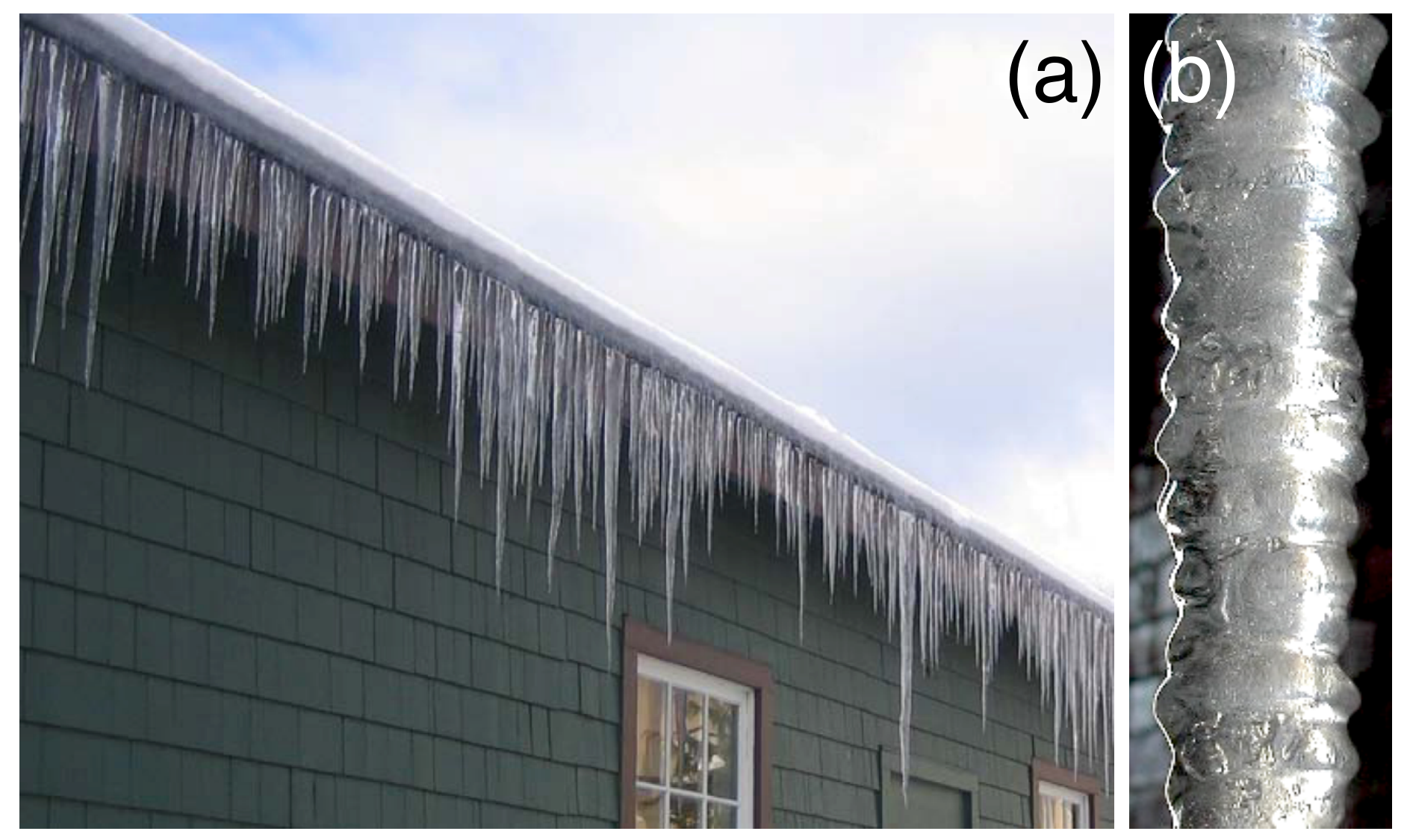}
\caption{ (Color online) (a) An array of icicles hanging from a rooftop on a cold Canadian winter day; (b) ripples on the surface of a natural icicle.}
\label{icicles}
\end{figure}

Icicles are harmless and picturesque features of cold winters in many parts of the world (Fig.~\ref{icicles}(a)). On the other hand, the unwanted build-up of ice on the surfaces of power lines~\cite{power} and airplane wings~\cite{airplanes} is a serious engineering hazard. In all cases, ice is formed when latent heat is transferred to the surrounding sub-freezing air from the growing surface. Typically, the liquid water is present only as a thin flowing film over the growing ice-water interface from which the heat is being transferred. On an icicle, this water film is subject to gravitational and viscous forces, as well as to surface tension forces at the water-air interface, and eventually collects in a pendant drop which periodically detaches from the tip.  An icicle grows by adding ice on its sides and, by a slightly different mechanism, by tip growth in the neighborhood of the pendant drop~\cite{makkonen1}.

While the kinematic growth mechanism of icicles is well-known~\cite{makkonen1}, predicting the emergent shape of an icicle is a non-trivial dynamic  free-boundary problem.  By coupling the transport of latent heat, the local physics of the water film, and the dynamics of the thermally buoyant boundary layer in the surrounding air, Short, \emph{et al.} made the remarkable prediction that icicles should have self-similar shapes which are related to those of stalactites~\cite{short1,short2}.   In this theory, axisymmetry is assumed and surface tension is neglected.  The freezing mechanism at the tip is not treated and tip growth is transformed away.  The self-similar shape that emerges is independent of details such as the water supply rate, the ambient temperature, and the thermodynamic parameters of water.  In this paper, we perform the first direct test of the self-similar theory of icicle shapes under controlled laboratory conditions.

The surface of an icicle can also become morphologically unstable, leading to the formation of ripple patterns~\cite{maeno,ogawa}, as shown in Fig.~\ref{icicles}(b).  This instability, which is presumably related to a similar effect in stalactites~\cite{ripply_stalactites}, crucially involves surface tension effects at the water-air interface~\cite{ueno1,ueno2,ueno3,ueno4,ueno5,rippleexpt2}, and is thus not captured by the aforementioned self-similarity theory for the overall shape.  In this paper, we describe observations of these ripple patterns in some laboratory-grown icicles and find that the motion of the ripples is upward during growth, in qualitative agreement with linear stability theory~\cite{ueno1,ueno2,ueno3,ueno4,ueno5,rippleexpt2}.

A complete theoretical understanding of the dynamics of icicle growth, incorporating shape, tip growth, and ripple formation, remains to be formulated.

This paper is organized as follows.  In the following section, we describe the icicle growing apparatus.  In section \ref{shape}, we discuss the detailed comparison of the self-similarity theory to the data, treating deviations from the self-similar shape in section \ref{nonideal}.  This is followed in section \ref{ripple} by a discussion of the motion of the ripple patterns observed.  Finally, section \ref{conclusion} presents a summary and conclusions.

\section{experiment}
\label{experiment}

\begin{figure}
\includegraphics[width=7.5cm]{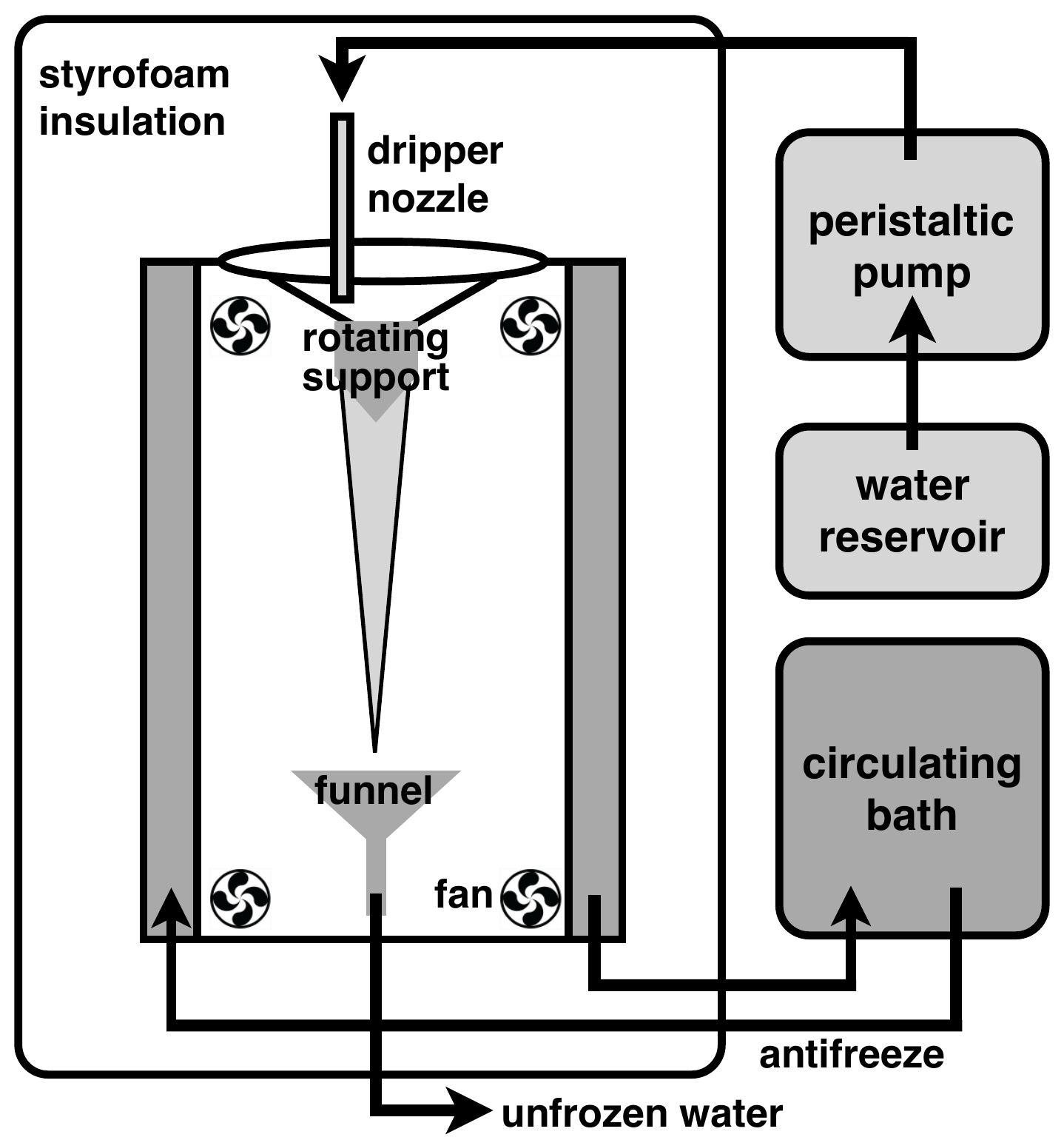}
\caption{ A schematic of the icicle growing apparatus.  }
\label{schematic}
\end{figure}

We conducted controlled icicle experiments using a table-top icicle growing apparatus, shown schematically in Fig.~\ref{schematic}.  Icicles were grown in an enclosed refrigerated box with dimensions  38 cm $\times$ 38 cm  $\times$ 107 cm in which air motion could be controlled with internal fans.  The temperature at the walls of the box was controlled to within $\pm0.3^\circ$C by a commercial bath which circulated antifreeze through pipes embedded in the walls.  The box was insulated with 10 cm of commercial foam insulation.   The icicle could be imaged with a computer controlled DSLR camera {\it via} a slot in the side of the box.  The optical path to the camera was insulated, and there were no windows between the camera and the icicle to avoid fogging. The icicle was illuminated by a distributed white LED light source against a black cloth background.

The growing icicle hung from a sharpened wooden support that was rotated to encourage axisymmetry and to allow all sides of the icicle to be imaged.  The rotation speed was typically 4 min per rotation, which was slow compared to air motions, but rapid compared to the $\sim$ 10 hr growth time of an icicle.  The rotational position of the support was indexed to allow the same view of the icicle to be imaged on successive rotations.  Water near 0$^\circ$C was supplied by a nozzle using a peristaltic pump.  The temperature of the nozzle was feedback controlled by the computer.  Rotation also assisted in the even distribution of water around the icicle.   The input nozzle was slightly offset from the rotation axis so that water reached the entire periphery of the support over one rotation.   Water that dripped off the icicle was collected by a warmed funnel.  Both tap water and distilled water were used.

Using this apparatus, we grew 93 icicles under varying conditions of air motion, water purity, input flow rate, and at ambient wall temperatures ranging between -7$^\circ$C to -21$^\circ$C.

\section{self-similar icicle shape}
\label{shape}

Using edge detection on images of growing icicles, we can directly test the self-similarity theory of Short \emph{et al.}~\cite{short1}.  This theory had previously been compared to eight images of natural icicles whose conditions of formation were not controlled~\cite{short1}.  In this section, we discuss our systematic comparison, and find good agreement with the self-similarity theory only in certain cases.  The sources of the various deviations from self-similarity are described in the following section.

By considering the latent heat transfer \emph{via} the water film and the rising thermal boundary layer~\cite{gold} near the icicle, Short, \emph{et al.}~\cite{short1} derived an ordinary differential equation for the scaled shape, 
\begin{equation}
\frac{d \rho}{d \zeta} = {1}\big/{\sqrt{\zeta^{\frac{1}{2}}-1}},
\label{one}
\end{equation}
where $\rho$ is the dimensionless radius of the icicle and $\zeta$ is the dimensionless vertical distance away from the tip.  Integrating Eq.~\ref{one} yields an equation for the predicted self-similar shape,
\begin{equation}
\rho(\zeta) = \frac{4}{3} \big( \zeta^{\frac{1}{2}} + 2 \big) \sqrt{ \zeta^{\frac{1}{2}} -1}.
\label{shapeeqn}
\end{equation}
This form is predicted to be an attractor in the sense that all icicles should converge to this shape as they grow.

To arrive at the dimensionless variables $\rho$ and $\zeta$, the physical dimensions of the icicle $r$ and $z$ must be nondimensionalized by a common factor $a$, which has dimensions of length.  According to the theory, 
\begin{equation}
a = a_{\rm th} = \ell \big( v_c / v_t )^4 = \frac{g \beta_a \Delta T^5}{\nu_a^2} \bigg[\frac{\Lambda_a}{v_t L C}  \bigg]^4 ~.
\label{a_expression}
\end{equation}
Here, $\ell$ is a length associated with the thickness of the rising boundary layer of warm air that surrounds the icicle.  $v_c$ is a characteristic speed that scales the speed of the slow surface growth of the icicle, while $v_t$ is the velocity of its growing tip.

A crucial feature of the theory is that $v_t$ is treated as an independent parameter and the physics of the tip growth is not further analyzed. The elongation of the icicle during growth is transformed away, so that only the geometry of the shape remains, with the tip treated as an infinitely sharp singularity.  The other parameters in Eqn.~\ref{a_expression} are the acceleration due to gravity $g$,  the volumetric expansion coefficient for air $\beta_a$,  the temperature difference between the icicle surface and the air far away $\Delta T$, the kinematic viscosity of air $\nu_a$, the thermal conductivity of air $\Lambda_a$, and the latent heat of fusion per unit volume of water $L$.  Finally, $C$ is a dimensionless constant of order unity which is related to the Prandtl number of air.

Using values of the parameters given by Short \emph{et al.}~\cite{short1}, $C=1$,  $\Delta T=12^\circ$C, and a typical observed  $v_t = 10^{-3}$~cm/s, we find that the theory predicts $a_{\rm th} \sim 2 \times 10^{-5}$~cm.  This length is much smaller than any geometric feature of icicles, and its physical interpretation is somewhat obscure.  For the purpose of this paper, $a$ functions simply as a theory-independent overall scale parameter in the fitting of fully-developed laboratory-grown icicles.



We used the \texttt{bwtraceboundary} function in the Matlab image processing toolbox to trace the edges in high resolution digital images of each icicle.  Then, using a Levenberg-Marquardt algorithm, we fit these icicle outlines to 
\begin{equation} \label{fullfunction}
\bigg[\frac{r}{a}\bigg] = \frac{4}{3} \bigg(\bigg[\frac{z}{a}\bigg]^{\frac{1}{2}}+2\bigg) \sqrt{ \bigg[\frac{z}{a}\bigg]^{\frac{1}{2}} -1}
\label{fitshape}
\end{equation}
to extract the scale factor $a$ that gives the least reduced $\chi^2$.  For each $z$, we defined $r(z)$ as half the distance between the two edges of the icicle; this allows slightly bent or tilted icicles to be analyzed. 

\begin{figure}
\includegraphics[width=7.2cm]{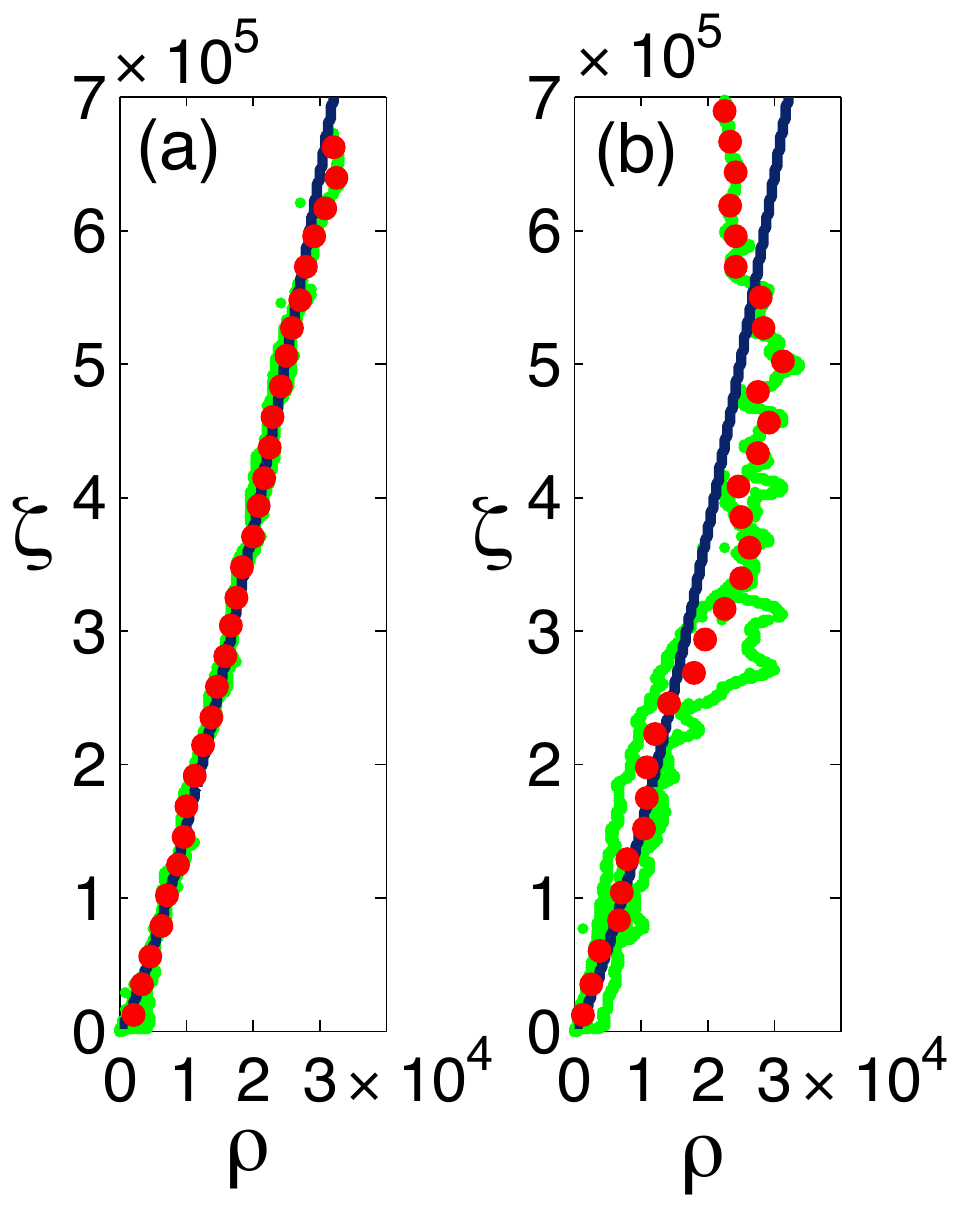}
\caption{ (Color online) Averaged plots of the dimensionless radius $\rho$ \emph{vs.} the dimensionless vertical distance $\zeta$ away from the tip for: (a) the five icicles with the smallest values of reduced $\chi^2$ from fits to the self-similar theory; (b) the five icicles with the largest values of reduced $\chi^2$.  The solid blue curve is the theoretical prediction, the small green circles in the background show the raw data, and the large red circles in the foreground show the averaged data.}
\label{bestvsworstfits}
\end{figure}

We were able to fit 76 icicles in this way.  The remaining 17 had multiple tips or other non-ideal features, as discussed in the following section.  In Fig.~\ref{bestvsworstfits}, we show data representative of the two extremes we observed among the icicles that were fit to the self-similar profile: Fig.~\ref{bestvsworstfits}(a) shows an averaged plot of the five fitted icicles with the smallest values of reduced $\chi^2$, and Fig.~\ref{bestvsworstfits}(b) shows an averaged plot of the five fitted icicles with the largest values of reduced $\chi^2$.

As can be seen from Fig.~\ref{bestvsworstfits}(a), {\it some} icicles do fit the self-similar profile to a very remarkable degree.  The good agreement in these cases is all the more impressive, because a variation of the single fitting parameter $a$ scales both axes equally, so that away from the tip, the slope of $\rho(\zeta)$ is extremely insensitive to $a$.  In all cases, we find that the fitted value of $a$ is extremely small, typically on the order of $10^{-3}\sim10^{-5}$~cm; this is in broad agreement with theoretical expectations.

\begin{figure}
\includegraphics[width=8.2cm]{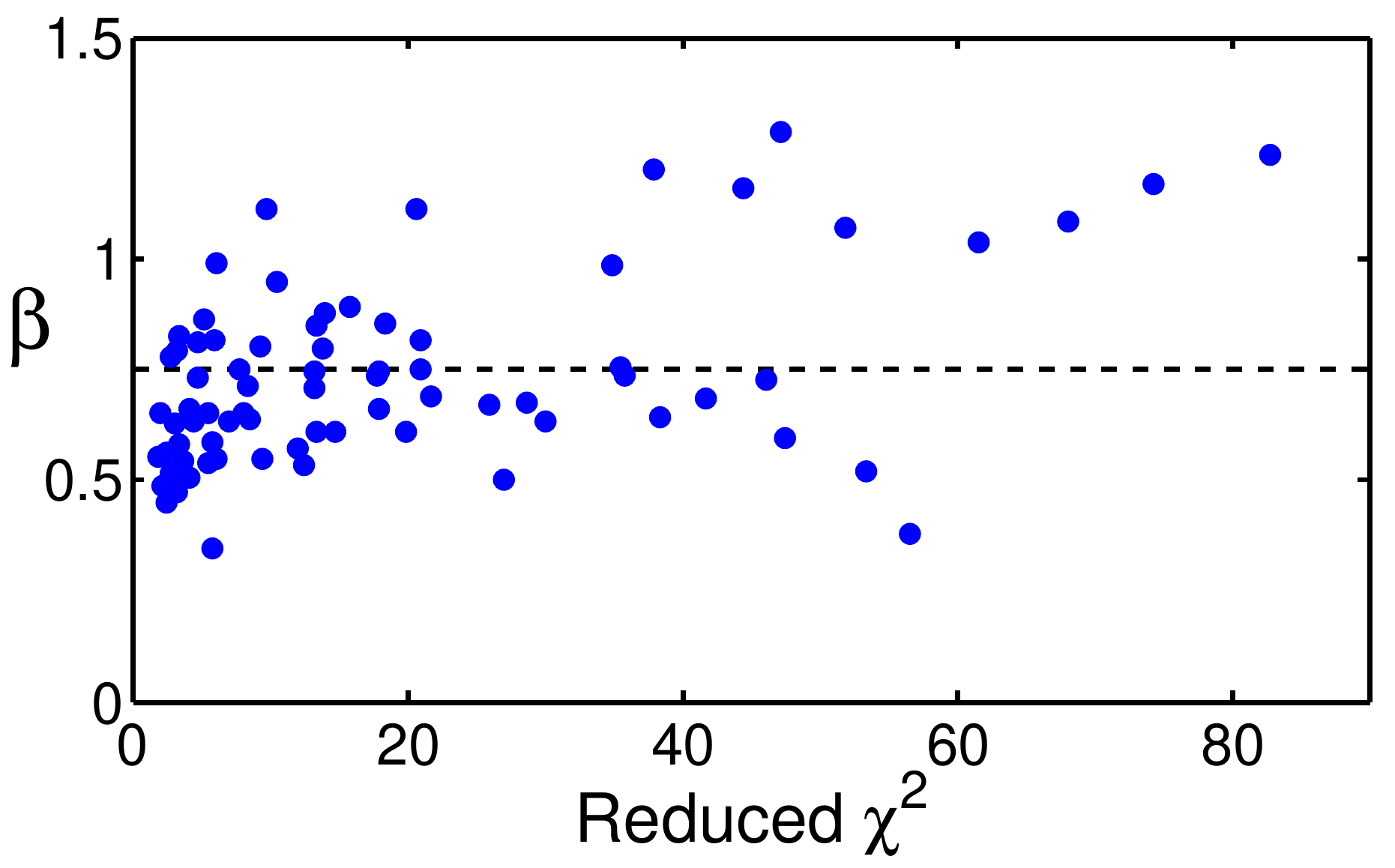}
\caption{(Color online) The best-fit exponent $\beta$ from fitting icicle profiles to the general power law $r=\alpha z^{\beta}$, plotted against reduced $\chi^2$.  The dashed line indicates the exponent $\beta=3/4$ expected from the self-similar theory.}
\label{powerlaw}
\end{figure}

Since $a$ is so small, for all practical purposes, the full functional form of Eqn.~\ref{shapeeqn} reduces to its small-$a$ limit, $\rho(\zeta) = (4/3) \zeta^{\frac{3}{4}}$.  The quality of the fits shown in Fig.~\ref{bestvsworstfits}(a) is a strong test of this specific 3/4 power law, as well as of its prefactor.  As a theory-independent model, we also tried a two-parameter fit of the unscaled icicle profile to an arbitrary power law $r=\alpha z^{\beta}$.  Fig.~\ref{powerlaw} plots the best-fit exponents $\beta$ against their corresponding reduced $\chi^2$.  Most, but not all, icicles that fit well to the self-similar theory scatter around $\beta=3/4$, indicated by the dashed line.  The pure power law, as an empirical two-parameter model, does about as well to describe the shape data as the full self-similar solution.

On the opposite extreme, Fig.~\ref{bestvsworstfits}(b) demonstrates that some icicles are rather badly described by the self-similar profile, or by any simple power law.  These icicles typically exhibit complex, non-monotonic shapes.  The agreement with the self-similarity profile is not convincingly improved by restricting the fit to the region of the tip. (Also recall that we have already excluded a number of icicles with even more pathological shapes, as discussed in the next section.)  Most icicles exhibit a degree of agreement with the self-similar profile that falls between the two extremes shown in Fig.~\ref{bestvsworstfits}.  

\begin{figure}
\includegraphics[width=8.2cm]{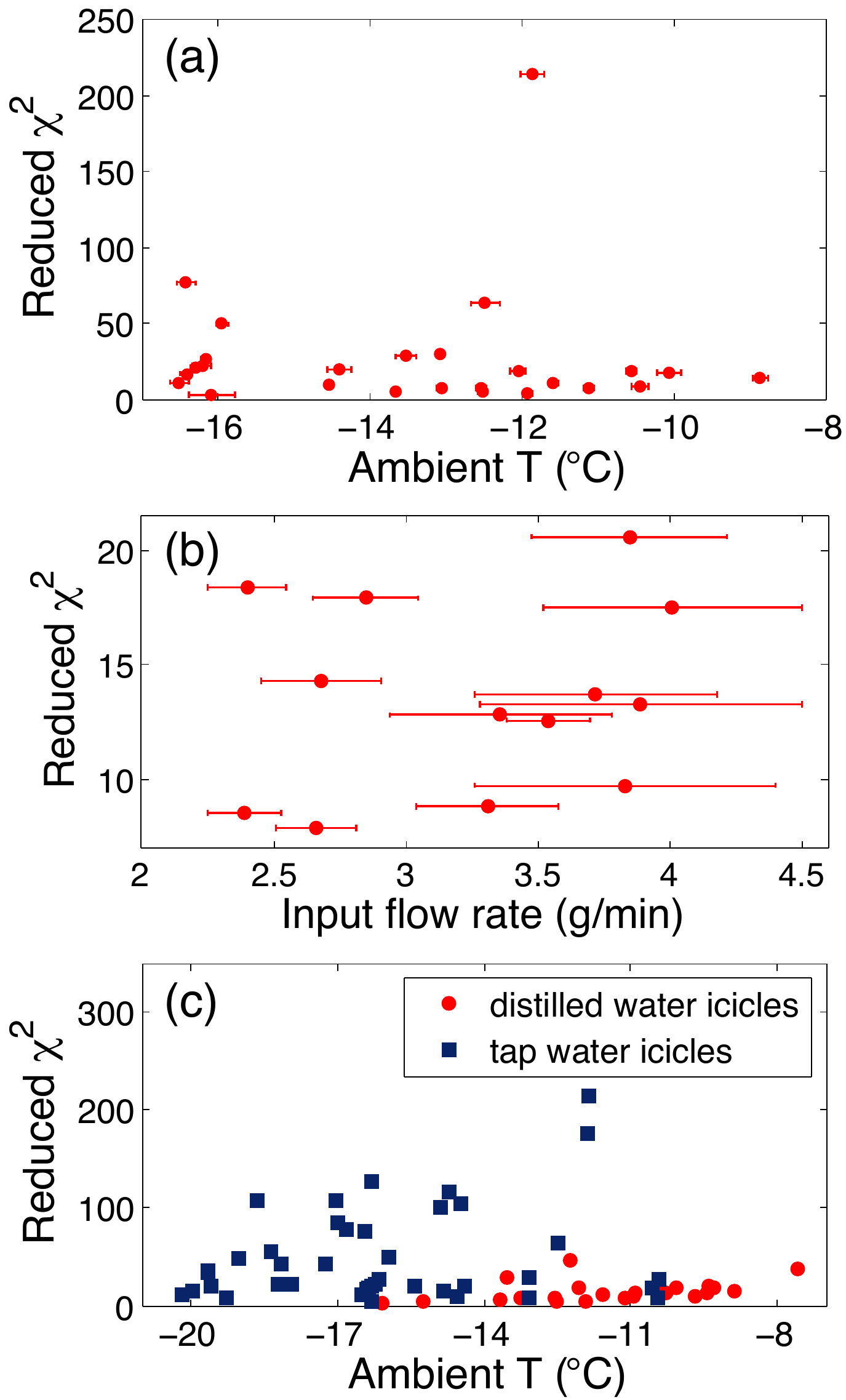}
\caption{(Color online) Values of reduced chi-square from fitting experimental icicle profiles to the self-similarity theory: (a) Reduced $\chi^2$ plotted against ambient temperature for 26 icicles grown at an input flow rate of 2.6$\pm$0.3 g/min.  (b) Reduced $\chi^2$ plotted against input flow rate for 13 icicles grown at an ambient temperature of -10$\pm$1$^\circ$C. (c) Reduced $\chi^2$ for 23 distilled water grown icicles and 41 tap water grown icicles; it is evident that distilled water grown icicles are better described by the self-similar shape.}
\label{reducedchisq}
\end{figure}

The quality of fit, as measured by the reduced $\chi^2$, shows no systematic dependence on the temperature of growth (Fig.~\ref{reducedchisq}(a)) or the water supply rate (Fig.~\ref{reducedchisq}(b)).  It does, however, show a clear dependence on the purity of the water used (Fig.~\ref{reducedchisq}(c)), with tap water grown icicles exhibiting a mean reduced $\chi^2$ of $\approx$ 50, {\it vs.} $\approx$ 14 for distilled water grown icicles.

\begin{figure}
\includegraphics[width=7.8cm]{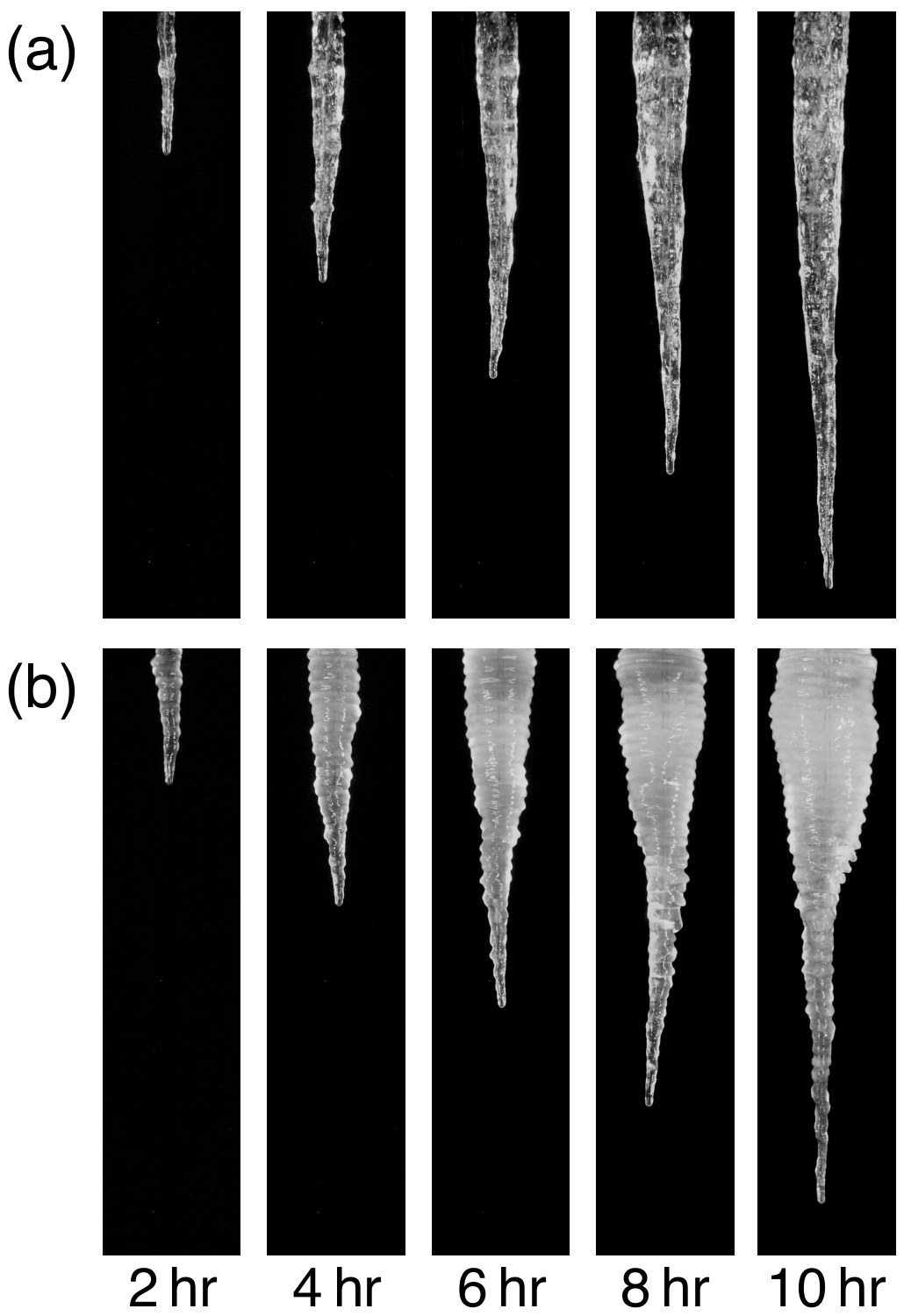}
\caption{ Photographic time-series of two icicles grown under identical conditions except for the purity of the water supplied: (a) distilled water {\it vs.} (b) tap water. The times given are the times elapsed since the initiation of the icicle growth.}
\label{distilledvstap}
\end{figure}

The morphological difference between icicles grown with tap water and distilled water is apparent to casual inspection.  Fig.~\ref{distilledvstap} shows time-lapse photographs of two icicles grown under identical conditions, with one made from tap water and the other from distilled water.  Tap water grown icicles have a bulge near their upper end and narrower tips than distilled water grown icicles, which are generally closer to the predicted monotonic self-similar shape.  Tap water grown icicles also exhibit much more prominent ripple patterns, as discussed in section \ref{ripple}.  Moreover, tap water icicles tend to be cloudier in appearance than distilled water grown icicles; this may reflect differences in the crystal structure of the ice and the presence of undissolved impurities in tap water that provide nucleation sites.

\section{non-ideal icicles and tip splitting}
\label{nonideal}

\begin{figure}
\includegraphics[width=8.5cm]{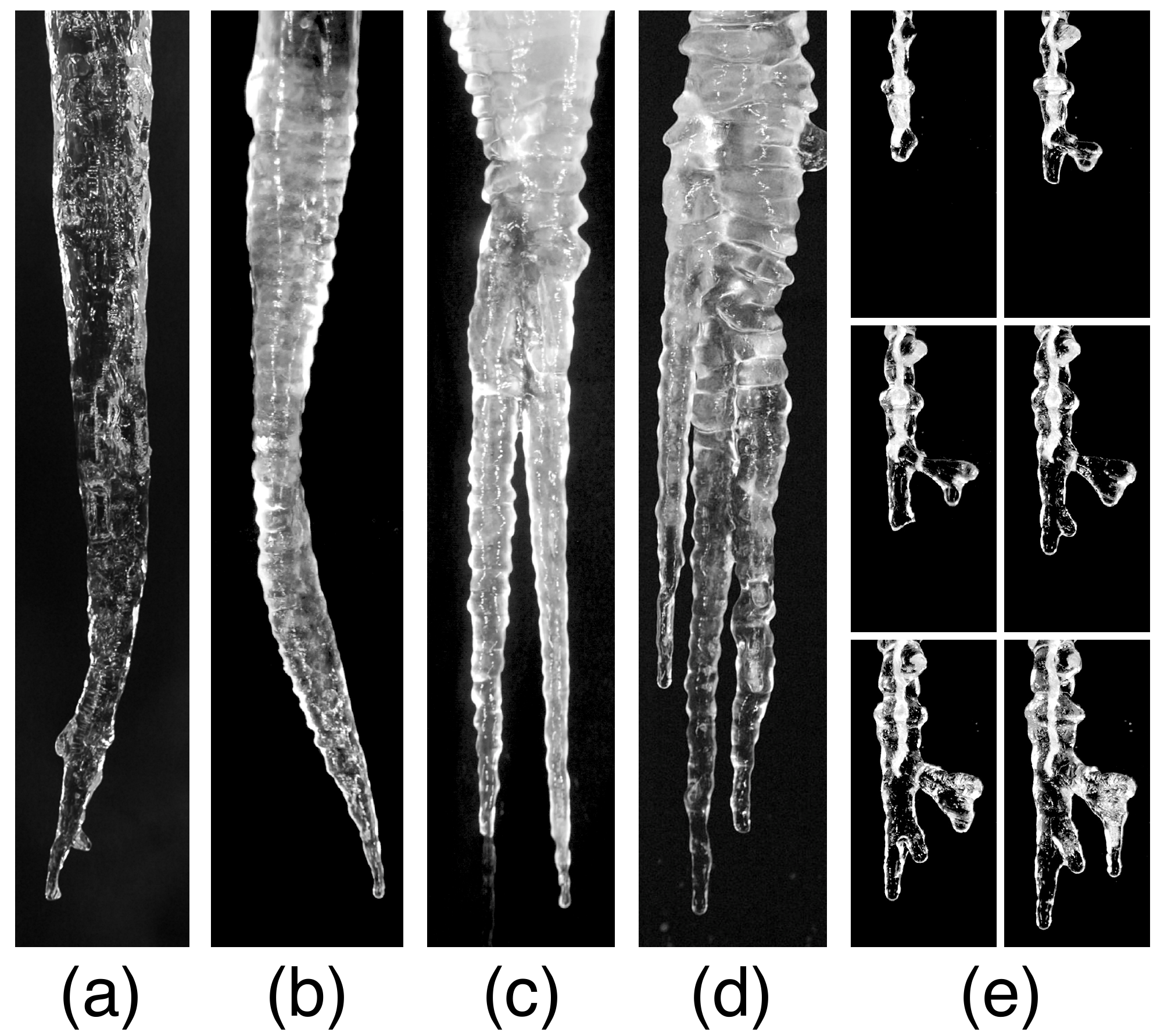}
\caption{ Some icicle non-uniformities: (a)-(b) bent, twisted icicles, (c)-(d) multitipped icicles, 
(e) time-series photographs of the initiation of a branched icicle grown on a piece of string (time interval: 10 minutes).}
\label{nonuni}
\end{figure}

A significant fraction of laboratory-grown icicles exhibit non-uniform shapes which are sufficiently pronounced that they cannot reasonably be described by the self-similar profile given by Eqn.~\ref{shapeeqn}.  Some of the most common shapes are shown in Fig.~\ref{nonuni}.    Some icicles bend or twist or otherwise strongly deviate from axisymmetry (Fig.~\ref{nonuni}(a)-(b)).  In some cases, these shapes may be due to strong wind effects arising from the fans used to circulate the air inside the refrigerated box.  In other cases, they may be due to uneven water distribution.  Even after considerable efforts were made to minimize these effects, such non-uniformities could not be completely eliminated.  This shows, at the least, that axisymmetry is not a strongly preferred state that emerges naturally from the growth process.

An unexpected but very commonly observed non-uniformity in laboratory grown icicles is branching of the tip, as shown in Fig.~\ref{nonuni}(c)-(d).  Fig.~\ref{nonuni}(e) shows the typical evolution pattern for the initiation of a branched icicle.  The branching process begins with the formation of a small protrusion at the surface of an icicle.  The protrusion usually appears a couple of cm above the tip and forms at some angle between 0$\degree$ and 90$\degree$ from the negative vertical axis.  Its length varies but is typically less than 4 cm.   Once a protrusion becomes a secondary source of pendant drops falling from the icicle,  it is inevitable that a branched icicle will result.  The various tips each continue to grow, sharing the incoming water supply.  Some icicles could still be fit to the self-similarity theory before branching occurred.

The presence or absence of air motion around the growing icicle was strongly correlated with the probability that branches will develop.   Of 20 runs conducted with the air-circulating fans off, 16 icicles (i.e.~80\%) branched at some point during their growth.   Of 73 icicles grown when the air was stirred, branching was observed in only 22 runs (i.e.~$\approx$30\%).  This is in very puzzling disagreement with the theory by Short, \emph{et al.}~\cite{short1}, which assumed a calm air environment and axisymmetry.  Instead, the most self-similar icicles were found when the air was stirred.

\section{ripple patterns}
\label{ripple}

Ripple patterns are commonly observed on natural icicles.  Such ripples are due to a morphological instability that is not contained in the self-similarity theory, but has been investigated by linear stability theory~\cite{ogawa,ripply_stalactites,ueno1,ueno2,ueno3,ueno4,ueno5} and experiments~\cite{rippleexpt1,rippleexpt2}.  The ripples are believed to involve surface tension effects which are neglected by the self-similarity theory.  According to the theory by Ueno, {\it et al.}~\cite{ripply_stalactites,ueno1,ueno2,ueno3,ueno4,ueno5}, increasing the surface tension of the water-air interface suppresses icicle ripples.  Moreover, they found ice grows faster just upstream of any protrusion and slower downstream, so they predicted the ripple pattern travels upwards during growth.

In the laboratory, ripples were commonly seen on tap water grown icicles, while most distilled water grown icicles were free of ripples.  A clear, regular ripple pattern appeared on 48 out of 53 icicles made with tap water.  On the other hand, only 2 out of 23 icicles made with distilled water had ripples.  The presence of ripples is presumably at least partially responsible for the larger values of reduced $\chi^2$ for tap water icicles, as shown in Fig.~\ref{reducedchisq}(c).  

We measured the surface tension of our water samples at 24.7$^\circ$C and found values slightly lower than the accepted surface tension of pure water of 0.07196 N/m, with those of distilled water tending to be the largest.  No sample, even of tap water, deviated by more than 3\% from the accepted value.  The trend between the probability of ripple observation and surface tension is thus qualitatively consistent with theoretical predictions~\cite{ripply_stalactites,ueno1,ueno2,ueno3,ueno4,ueno5}, although it seems remarkable that surface tension differences of only a few percent could have such large effects on ripple formation.

\begin{figure}
\includegraphics[width=8.8cm]{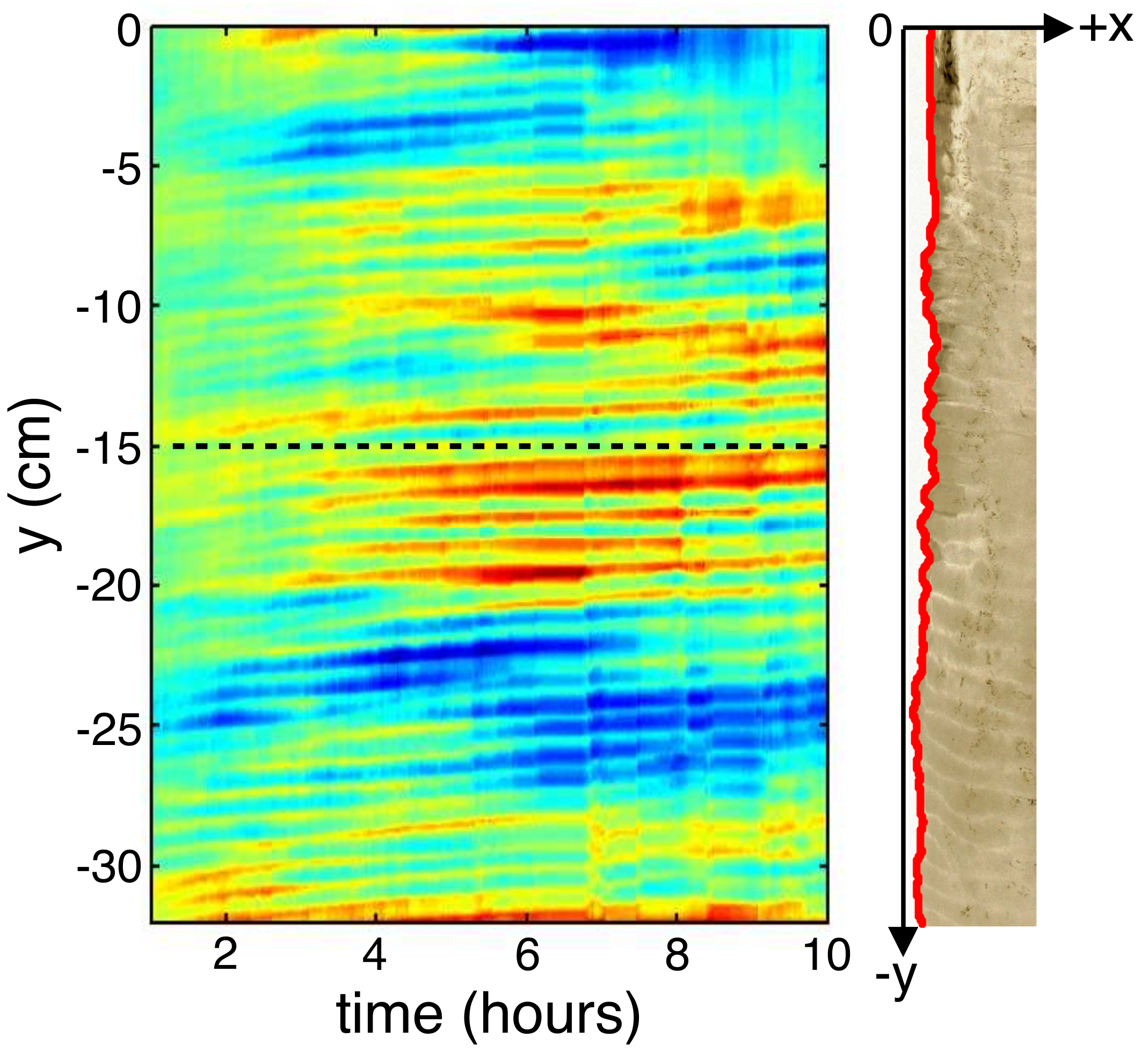}
\caption{ (Color online) A space-time plot of ripples on a long vertical finger of ice, using the coordinate system shown on the right.  The $x$ coordinate of the edge is plotted as a function of the $y$ coordinate and time.  The dashed horizontal reference line shows that the ripples moved slightly upward during growth.}
\label{ripplemotion}
\end{figure}

A particularly good sample of ripples was found on a long finger of ice that fortuitously formed when the drain of the refrigerated box froze.  The cylindrical finger grew from the bottom upward and bridged the whole height of the apparatus.  The rotation of the icicle support was turned off.  As the tap water continued to flow down the exterior of the finger, an extensive pattern of ripples was formed, which could be studied independently of the overall icicle shape.  These ripples are shown in Fig.~\ref{ripplemotion}.  We found that the ripples moved up the ice surface during growth, in qualitative agreement with the prediction of linear stability theory of Ueno, {\it et al.}~\cite{ripply_stalactites,ueno1,ueno2,ueno3,ueno4,ueno5}.  The wavelength we found agrees with that previously observed~\cite{rippleexpt1,rippleexpt2}.

\section{conclusion}
\label{conclusion}

In summary, we constructed an apparatus for the controlled growth of icicles.  The apparatus allows for the systematic study of icicle shapes under different conditions of temperature, input water flow rate, and ambient air motion.  It uses a rotating support and other features to encourage axisymmetry.  Using this apparatus, we grew 93 icicles under various conditions and compared their shapes to the self-similar theory of Short, \emph{et al.}~\cite{short1}.

While we found very close agreement in some cases, a large fraction of the icicles grown deviated significantly from self-similarity.  The degree of deviation was not systematically correlated with the temperature conditions of growth or the water supply rate, but was rather dependent on the purity of the water used and the movement, or lack of it, in the surrounding air.  The most ideal icicles were found for distilled water and gently stirred air.  The latter condition contradicts the assumptions of the self-similarity theory, but nevertheless improves the agreement with it.  Icicles grown in still air had a higher probability of forming multiple tips.

We also examined the ripple patterns on the surface of icicles.  We found that ripples were suppressed on the most ideal icicles.  In cases where the formation and growth of ripples were clear, they were observed to climb the icicle during growth, in qualitative agreement with the linear stability theory of Ueno, \emph{et al.}~\cite{ripply_stalactites,ueno1,ueno2,ueno3,ueno4,ueno5}.  A detailed study of ripple patterns will require more controlled experiments on cylindrical substrates and possibly the addition of surfactants to vary the surface tension.  A complete theory of icicle morphology, including tip growth, self-similarity, and the rippling instability, is currently lacking.

\begin{acknowledgments}
We thank Charles Ward for his help with the surface tension measurements.  This research was supported by the Natural Science and Engineering Research Council (NSERC) of Canada.
\end{acknowledgments}

\end{document}